\documentclass[author,type1rest]{nrc1}
\journal{Ca. J. Phys.}
\usepackage{graphicx}
\usepackage[figuresright]{rotating}
\usepackage{cite}
\begin{document}
\title{Oscillator strength measurements of atomic absorption lines from stellar spectra\thanks{Based in part on observations made with the Mercator Telescope, operated on the island of La Palma by the Flemish Community, at the Spanish Observatorio del Roque de los Muchachos of the Instituto de Astrof\'{i}sica de Canarias.}}
\author{A. Lobel}
\correspond{Alex.Lobel@oma.be; alobel@sdf.lonestar.org}
\address{Royal Observatory of Belgium, Ringlaan 3, B-1180, Brussels, Belgium}
\shortauthor{A. Lobel}
\maketitle
\begin{abstract}
We develop a new method to determine oscillator
strength values of atomic absorption lines with state-of-the-art
detailed spectral synthesis calculations of the optical
spectrum of the Sun and of standard spectral reference stars.
We update the log(gf)-values of 911 neutral lines observed
in the KPNO-FTS flux spectrum of the Sun and
high-resolution echelle spectra ($R$=80,000) of Procyon (F5 IV-V)
and $\epsilon$ Eri (K2 V) observed with large signal-to-noise (S/N) 
ratios of $\sim$2,000 using the new Mercator-Hermes 
spectrograph at La Palma Observatory (Spain).

We find for 483 Fe~{\sc i}, 85 Ni~{\sc i}, and 51 Si~{\sc i} 
absorption lines in the sample a systematic over-estimation of the 
literature log(gf)-values with central line depths below 15 \%. 
We employ a curve-of-growth analysis technique to test the 
accuracy of the new oscillator strength values and 
compare calculated equivalent line widths to the 
Moore, Minnaert, \& Houtgast atlas of the Sun.

The online SpectroWeb database at {\tt spectra.freeshell.org} 
interactively displays the observed and synthetic spectra 
and provides the new log(gf)-values together with important 
atomic line data. The graphical database is
under development for stellar reference spectra of every spectral
sub-class observed with large spectral resolution and 
S/N ratios.
\end{abstract}

\section{Introduction}
Accurate oscillator strength values are of central 
importance for the analysis and interpretation of astrophysical 
spectra. Detailed radiative transfer modeling of the profiles 
of absorption and emission lines observed in the spectrum 
of the Sun and stars requires reliable atomic data to 
correctly infer the physical properties of the atmospheric 
line formation region. These data are however currently unavailable 
for a large fraction of the weak and medium-strong absorption 
lines observed in optical stellar spectra. In recent years 
a number of online databases have become available that 
offer atomic line data compiled from a large variety 
of sources in the scientific literature, such as {\sc nist}, 
{\sc vald}, the Kurucz website, Topbase, etc. Thanks to the
efforts of the atomic database developers the detailed 
synthesis of stellar spectra has become much more
accessible to astrophysicists without having to fully address 
the fundamental atomic physics needed to acquire
these important atomic input data. On the other hand, 
the successful application of publicly available atomic databases 
entirely depends on the quality of the data compiled 
in them, but which is frequently not (or more appropriately
cannot always be) guaranteed.

The demand for spectral standard star atlases is 
steadily increasing with the fast improvements of 
resolving power and the quality by which stars of 
all spectral types are observed with modern spectrographs. 
Printed atlases of stellar spectra often only provide 
a small list of identified features without an assessment 
of the reliability of the spectral line identifications. 
Users often have no means to determine if the spectral line 
identifications are valid, or if they have been revised since 
publication. On the other hand, many atomic databases offer
line data (that can be text queried online) 
based in part on theoretical calculations that have not 
been tested against observed stellar spectra. Users 
cannot readily verify the quality of the line data, 
or if it applies to their spectroscopic observations. 
The data can contain line identifications that do not 
apply to an observed stellar spectrum because of unknown 
atmospheric formation conditions or elemental abundance 
differences with the solar values. Conversely, observed 
spectral features often cannot be identified because 
the quality of the provided atomic and molecular line 
data is limited and requires further improvements.

In this paper we present a critical evaluation of 911 oscillator 
strength values of astrophysical importance based on 
detailed spectral synthesis calculations of the 
weak and medium-strong neutral lines spectrum of the Sun 
between 400 nm and 680 nm, and of two bright spectroscopic reference 
stars, Procyon and $\epsilon$ Eri, observed with large 
spectral dispersion, and with S/N ratios exceeding 2,000.
The observed and best-fit theoretical spectra are publicly 
available in the online 
SpectroWeb\footnote{\tt alobel.freeshell.org/spectrowebl.html} 
database \cite{lobe1}. 
Users can directly assess the quality of the absorption line 
identifications by comparing the observed spectra 
with state-of-the-art computed spectra. The graphical web 
interface enables users to select 10 or 25 \AA\, wide 
spectral regions-of-interest from an interactive list of 
observed wavelengths, together with the employed atomic line 
data. We investigate the accuracy of the log(gf)-values 
obtained with best fits to the observed spectra using an 
average curve-of growth analysis method. We also discuss a 
comparison of computed and observed line equivalent widths
to investigate remarkable trends in a number of important neutral
elements that reveal systematic over-estimations of 
literature (e.g., that are offered in online atomic 
databases) log(gf)-values for weak Fe~{\sc i}, 
Ni~{\sc i}, and Si~{\sc i} lines with central 
line depths below 15 \%.

\section{Observations}

We observe the optical spectra of the white F5-type main sequence 
star Procyon ($\alpha$ CMi A; HD~61421) and the narrow-lined K2-type 
dwarf $\epsilon$ Eri (HD~22049) with the new Hermes spectrometer 
on the 1.2 m Mercator telescope at La Palma Observatory, Canary Islands. 
The instrument is a prism-cross-dispersed fiber-fed 
bench-mounted echelle spectrograph that observes the complete 
wavelength range from 377 nm to 900 nm in a single exposure. 
In high-resolution mode the spectral resolving power is $R$=80,000
using a 2.5 arcsec fiber equipped with an image slicer and 
optimized for very high efficiency of $>$ 25\% in 
the optical band. For Procyon ($V=0^{\rm m}.34$) we 
achieve a combined S/N ratio$\simeq$2,200 with 50 consecutive 
exposures of 20 seconds, while for $\epsilon$ Eri ($V=3^{\rm m}.73$) 
the S/N ratio$\simeq$2,000 with 20 exposures of 160 seconds. 

The spectra are calibrated with a data reduction pipeline 
developed at the ROB. The typical echelle pipeline calibration
steps are performed starting with bias subtraction, 
optimal order extraction, division of the science exposures 
with an averaged flat-field frame, and followed by the 
order wavelength calibration using a Th-Ar lamp spectrum.       
The flat-fielding step also removes the blaze-function before
co-addition of all exposures. It yields a final 1-D 
order-merged spectrum with a rather characteristic 
low-frequency shape we utilize for an initial estimate of the 
complete continuum flux normalized spectrum.
Careful design including efficient baffling and 
use of high-quality optical components, allow Hermes to deliver 
very clean spectra, comparing favorably with many other
echelle spectrographs. The distribution of scattered light 
in the spectrograph is very local. The inter-order background 
signal fluctuates around 0.1~\% of the flux in the adjacent orders
\cite{rask1}.

The high-resolution spectrum of the Sun observed in 1981 with the 
NSO/KPNO Fourier Transform Spectrograph (FTS) at the McMath-Pierce 
Solar Facility is offered in \cite{neck1}. The spectral 
resolution of the FTS flux spectrum is 
$R$$\sim$350,000, with an estimated S/N ratio$\sim$2,500. 
More information about the calibration of the solar intensity 
(averaged over the solar disk) atlas is provided in \cite{neck2}.

\section{Continuum normalization and spectrum synthesis calculations}
 
An important problem for modeling absorption lines in 
large S/N ratio spectra is the flux normalization to the 
stellar continuum flux level. 
The placement of continuum is crucial for accurate 
improvements of atomic line data with detailed synthesis 
calculations. Firstly, we apply a standard continuum 
normalization procedure to the Hermes spectra. 
The initial normalization step is fully automated 
because the blaze-function is very precisely removed. 
It searches for continuum flux reference points in pre-selected 
wavelength regions void of broad stellar lines and 
telluric absorption features, and computes the overall shape 
of the spectrum using a mean weighted flux method. 
The normalized fluxes are generally very accurate. 
In the second normalization step the observed spectra are 
correlated with synthetic spectra which improves
the accuracy of the placement of the continuum level to 
typically $\sim$1\%-2\%.

The synthetic spectra
are calculated with the LTE radiative transfer
{\sc scanspec}\footnote{\tt alobel.freeshell.org/scan.html} code. 
The code is based on the older {\sc scan} code 
(\cite{nieu1}), extensively used for studies of microturbulence 
velocities in massive cool stars. It iteratively solves 
the Milne-Eddington LTE transfer equation in 1-D stellar 
atmosphere models\cite{kuru1}. Important line broadening
effects for strong resonance lines and the stellar continua are
treated, which also incorporates an opacity distribution function.
The equation of state includes the most important hydrides, 
$\rm C_{2}$, CN, CO, and a number of oxides with updated 
partition functions. Hyperfine splitting in Mn~{\sc i}, 
Cu~{\sc i}, Co~{\sc i}, and in lines of some other elements 
is currently not treated. The line-of-sight 
microturbulence velocity $\zeta_{\mu}$ is assumed 
to be constant with depth in the atmosphere model.
Extensive statisitcal studies with {\sc scan} 
of the formation of optical Fe~{\sc i} and Fe~{\sc ii} lines 
observed in Canopus \cite{achm1} and in $\eta$ Leo 
\cite{loba1} show that the  
microturbulence velocities are practically constant 
with depth in 1-D atmosphere models of these stars.

We use atmosphere model parameters for the Sun
of ($T_{\rm eff}$ in K, log $g$ in $\rm cm\,s^{-2}$, 
$\zeta_{\mu}$ in $\rm km\,s^{-1}$) 
= (5777, 4.438, 1.1). For Procyon we use 
(6550, 4.0, 1.2), and  (5050, 4.5, 0.55) for $\epsilon$ Eri.
The microturbulence velocities are varied until an 
overall best fit to the observed spectra is obtained. 
The synthetic spectra are convolved with the geometric 
mean of the $v$sin$i$ and macroturbulence velocity values.
2.5 $\rm km\,s^{-1}$ is used for the Sun, 3.6 $\rm km\,s^{-1}$ 
for Procyon, and 2.3 $\rm km\,s^{-1}$
for $\epsilon$ Eri. Notice the small broadening in 
$\epsilon$ Eri (i.e., \cite{take1} finds $v$sin$i$=2~$\rm km\,s^{-1}$) 
yielding the very narrow absorption lines in this orange-red 
dwarf $\sim$700 K cooler than the Sun. The theoretical spectra 
are convolved with the appropriate filter functions that 
simulate the instrumental resolving power.

We adopt the line broadening parameters (the four line damping 
constants) in the {\sc vald-2}\cite{kupk1} database.
The input values of log(gf), the line rest wavelengths, and
the transition energy levels are adopted either from 
{\sc nist}, {\sc vald-2}, or the Kurucz website, providing
the best initial fit to the observed absorption line positions 
and depths. A large number of diatomic molecular lines of
 \cite{kuru2} is also 
included to improve the computed fluxes due to weak 
molecular background absorption we clearly observe 
in $\epsilon$ Eri. The spectra are currently computed 
for solar elemental abundance values with atmosphere 
models of [M/H]=0.0. We use surface abundance 
values of \cite{ande1} for consistency with the 
models. For example, we adopt an iron abundance 
of 7.67, instead of the more recent and appreciably 
smaller (meteoritic) value of 7.5 \cite{grev1}. 
Procyon and $\epsilon$ Eri have published metallicities 
essentially equal to the solar values. \cite{heit1} 
finds [Fe/H]=$-$0.01$\pm$0.08 for $\epsilon$ 
Eri (see also \cite{kato1}). \cite{luck1} provides 
[Fe/H]=$-$0.04$\pm$0.06. \cite{kato1} derives 
for Procyon an average iron abundance from Fe~{\sc i} lines
identical to the solar value. More recently, 
\cite{kerv1} applied an iron surface abundance of 
[Fe/H]=$-$0.05$\pm$0.03.

Variability of the spectral line depths or shapes 
is not observed in these optically bright dwarfs. 
Observations with the Hipparcos satellite 
in 1990-1993 reveal only small $V$ variability with 
$\sigma V_{\rm Hip}$ $<$$0^{\rm m}.0017$ 
for Procyon, and $<$$0^{\rm m}.0014$ for $\epsilon$ Eri (see \cite{vanL1}).
\cite{brun1} find a granulation signal in Procyon 
observed with the WIRE satellite over $\sim$10 days. 
Variability in the visual brightness curves with 
very small amplitudes of 8.5$\pm$2 ppm 
are found to be consistent with possible stellar 
p-mode oscillations. \cite{frey1} 
observe $\Delta$$V$=$0^{\rm m}.01$-$0^{\rm m}.03$ over 
10 to 12 days attributed to the rotation of small 
surface starspots on $\epsilon$ Eri.

\section{Log(gf) measurements in the Sun, Procyon, and $\epsilon$ Eri}

We show in Fig. 1 a comparison of computed and observed spectra 
of $\epsilon$ Eri ({\it top panels}), the Sun ({\it middle panels}),
and Procyon ({\it bottom panels}) for 1.8~\AA\, around 5619~\AA.
The observed spectra ({\it solid black lines}) show two 
weak ($\lambda_{0}$=5619.174~\AA\, and 5620.027~\AA), and two 
medium-strong Fe~{\sc i} absorption lines ($\lambda_{0}$=5618.631~\AA\, 
and 5619.590~\AA). The line depths and shapes of 
three lines in the theoretical spectra ({\it solid blue lines}) 
correctly fit (e.g., within 1\% to 2\%) the observed spectra of 
the three stars using literature (viz. atomic database) log(gf)-values. 
The weak Fe~{\sc i} line at 5620.027~\AA, however, 
does not fit the observed line depths. 
The line with adopted log(gf)=$-$2.324 is computed $\simeq$10\% 
too strong in $\epsilon$ Eri, $\simeq$9\% in the Sun, 
and $\simeq$3\% in Procyon. We iteratively adapt the log(gf)-value
until we find a best fit to the depth of the line observed 
in the three spectra. We obtain the best fit 
for log(gf)=$-$2.724 ({\it solid red lines}), or by 
decreasing the initial log(gf)-value with 
$\Delta$log(gf)=$-$0.41. To limit long 
iteration times the minimum step size 
is set fixed to 0.1 dex, except for the final iteration 
step size which is determined by linear interpolation.
For comparison we also show the line computed in the
solar spectrum with log(gf)=$-$2.724$\pm$0.1 ({\it thin 
drawn solid red lines in the middle panels}).

We perform the iterative log(gf)-adjustment procedure  
for 911 neutral lines observed between 400 nm and 680 nm.
The measured lines are almost uniformally distributed 
over this range of wavelengths. Figures 2 \& 3 of \cite{lobe2} 
show a somewhat smaller number of medium-strong Fe~{\sc i} 
lines we adjust longward of 600 nm because the total 
number of medium-strong lines also decreases over this 
wavelength range.We select lines with clearly observed 
flux minima, if possible rejecting strong blends.
We do not select strong lines on the root part of the curve of growth
(c-o-g). All selected lines are on the linear part and a small number
are on the lower portion of the flat part of the c-o-g (Sect. 5).
We reject candidate lines with nearby telluric features or 
that reveal obvious contamination.

We find a remarkable systematic trend in the 
$\Delta$log(gf)-values (hereafter called `$\Delta$s') in Fig. 2. 
The major fraction of weak lines in the sample of 911 
neutral lines have adopted log(gf)-values we must 
substantially decrease to correctly 
fit the line depths observed in the three stars. 
The log(gf)-values we measure for 483 Fe~{\sc i} lines 
({\it open red circles}) are on average below the diagonal, 
and the $\Delta$s become systematically more negative 
towards smaller log(gf)-values below $-$2. The log(gf)-values we 
measure for 51 Si~{\sc i} lines ({\it open blue triangle 
symbols}) and for 85 Ni~{\sc i} lines ({\it open green 
square symbols}) are on average smaller than the
literature log(gf)-values. However, towards the stronger lines 
(e.g., log(gf)$>$ $-$2) we measure log(gf)-values that
on average exceed the literature values, or the medium-strong
lines of the sample require systematically more 
positive $\Delta$-values ({\it solid black symbols}).

Figure 3 shows a plot of the $\Delta$s with the (computed) 
central depths of the atomic lines we adjust 
in the three stars. Note that the continuum normalized 
line depths are computed without instrumental broadening. 
We observe systematic trends of more negative $\Delta$s 
towards weaker Fe~{\sc i} ({\it upper left-hand panel}) 
and Ni~{\sc i} lines ({\it lower middle panel}). The trends 
are more obvious in $\epsilon$ Eri ({\it open blue triangle symbols})
than in the Sun ({\it solid black symbols}) (and Procyon) 
because the neutral lines spectrum is stronger in 
cooler stars. Smaller systematic trends are also observed 
in lines of two other important iron-peak elements Ti~{\sc i} (82 lines)
({\it upper middle panel}) and Cr~{\sc i} (75 lines) 
({\it lower left-hand panel}). However, we do not 
find a similar systematic trend in 41 measured V~{\sc i} lines 
({\it lower right-hand panel}) of our sample.
We investigate the lines of 13 other
neutral elements but do not statistically significantly
detect systematic trends, except for 47
Co~{\sc i} lines for which we measure a rather weak
trend.  

In the next section we investigate possible causes for 
the systematic trends in Fe~{\sc i} and Si~{\sc i} lines 
for which we adjust the log(gf)-values adopted from 
atomic databases. For this purpose we compare 
computed and observed (e.g., published) equivalent line 
widths. We show that the remarkable trends cannot be 
attributed to large systematic errors in the spectrum 
synthesis calculations.

\section{Observed equivalent line widths and curve of growth analyses}

We investigate the systematic trends in the $\Delta$s of
Fe~{\sc i} and Si~{\sc i} lines by comparing with the
equivalent line widths (Weq) observed in the Sun by 
\cite{moor1} (hereafter MMH). The MMH atlas offers 
observed Weq-values (and reduced widths) 
of $\sim$24,000 (revised) identified solar 
absorption lines the authors systematically measure with a procedure that 
can also disentangle blended weak lines. The Weq-values
of 312 identified Fe~{\sc i} lines offered in the MMH 
atlas are calculated with the log(gf)-values we measure 
in Sect.~4. Figure 4 plots the $\Delta$s 
with the central depths of 483 lines ({\it upper left-hand panel}), 
and with the Weq-values computed below 150 m\AA\, for 
these 312 lines ({\it upper right-hand panel}). 
We find a strong correlation of the computed and 
observed Fe~{\sc i} Weq-values ({\it lower 
right-hand panel}). The lower left-hand panel shows the 
$\Delta$s compared to the differences between the observed 
and computed equivalent line widths ($\Delta$Weq) for
Weq $<$ 60~m\AA. It reveals that there are no systematic
dependences of the $\Delta$s on the $\Delta$Weq-values. 
We find that the ($\Delta$Weq, $\Delta$)-values are nearly 
symmetrically distributed around the center-of-gravity 
of the points cloud. The mean difference between observed 
and computed Weq-values is $\sim$5 to 10~m\AA, with 
maximum $\Delta$Weq-values below $\sim$20 m\AA\, for 
the strongest lines of the sample. The near-symmetry 
of the ($\Delta$Weq, $\Delta$)-values
shows that the mean errors of the $\Delta$s for weak lines 
are essentially randomly distributed. 
Possible systematic errors in our spectrum synthesis 
calculations therefore must remain sufficiently 
small compared to the random errors. Note that such 
random errors in Weq measurements are often due to inaccurate
placements of the local continuum flux level, scattered
(stray) light in the spectrograph not precisely removed 
during spectrum calibration, poor multi-Gaussian fits
to overlap of the line wings, unaccounted
hidden line blending i.e. from weak 
unidentified molecular absorption features, etc. 

A comparable analysis of 48 Ni~{\sc i} and 31 Si~{\sc i} 
solar lines in the MMH atlas reveals a similar near-symmetry 
of the ($\Delta$Weq, $\Delta$)-distributions.
It signals that the sources for the systematic trends 
we measure in these elements are present in the 
log(gf)-values we adopt from the atomic databases in Sect. 4. 
We test this hypothesis by calculating  
the standard mean errors of the log(gf)-values we 
measure from the observed spectra 
based on an average curve-of-growth method.

The upper panels of Fig. 5 show the $\Delta$s of 483 Fe~{\sc i}
lines ({\it left-hand panel}) and of 51 Si~{\sc i} lines
({\it right-hand panel}) with the reduced equivalent line width
log(Weq/$\lambda$) we compute in the Sun. The middle
panels show the corresponding curves-of-growth computed
with the log(gf)-values we measure for all lines of 
each species. The measured log(gf)-values are subtracted by 
$\theta$$\times$$\chi_{\rm low}$, where $\chi_{\rm low}$
is the lower excitation energy level (in eV), and 
$\theta$=5040 K / $T_{\tau_{5000}=1}$ is the inverse temperature. 
The curves-of-growth of both elements are sharply 
defined and can be utilized to compute relative
errors from the mean of measured log(gf)-values.
We compute the mean (curves-of-growth) with a least-squares 
$4^{\rm th}$-order polynomial regression fit to the
log(Weq/$\lambda$)-values ({\it solid red lines}). We find 
that medium-strong Fe~{\sc i} lines with log(gf)$\geq$$-2$ 
are on the flat part of the curve-of-growth. On the other
hand, all the Si~{\sc i} lines we measure are on the linear part 
of the curve-of-growth ({\it lower right-hand panel}). 

For the medium-strong Fe~{\sc i}
lines we compute maximum residual errors (relative to  
the mean curve-of-growth) that are typically a few $\sigma$-values. 
The residual errors of the measured log(gf)-values are marked 
with the errorbars in the upper panels. Weak Fe~{\sc i} lines 
with log(gf)$\leq$$-2$ are on the linear part of the 
curve-of-growth and reveal typical residual errors below 
one-$\sigma$ value of 0.05 to 0.1 dex (they are often smaller 
than the size of the marked symbols in the upper panel). 
Hence, the residual error values of the $\Delta$s of 
the weak lines are typically smaller than the minimum step 
size of 0.1 dex we apply to iteratively adjust 
(with a final interpolated step size below 0.1 dex) 
the log(gf)-values we adopt from atomic databases. 
It signals that our semi-empirically determined 
log(gf)-values are accurately measured using the 
spectral line synthesis iteration method with 
three stars outlined in Sect. 4.

We compute similar curves-of-growth of all other
important neutral elements for which we measure 
the log(gf)-values in the Sun, $\epsilon$ Eri, and Procyon. 
The residual errors of the log(gf)-values of 
weak lines computed from an average curve-of growth 
analysis are typically below one-$\sigma$ value ($<$0.1 dex). 
It strongly indicates that the systematic trends we 
find for the weak Fe~{\sc i}, Ni~{\sc i}, and Si~{\sc i} lines  
cannot be attributed to large systematic errors in 
the spectral line synthesis measurement method.
The small relative errors of the measured log(gf)-values 
point to systematic trends in the log(gf)-values we adopt 
from atomic databases for weak lines with normalized 
central depths below 15 \%.

We also plot for comparison in the lower panels of Fig. 5 
the curves-of-growth based on {\em observed} Weq-values
of 312 Fe~{\sc i} ({\em left-hand panel}) and 
31 Si~{\sc i} ({\em right-hand panel}) lines 
identified and measured by MMH in the solar spectrum. 
The standard deviation ($\sigma$) of the log(gf)-values we 
measure with respect to the mean curves-of-growth
({\em solid red lines}) are a factor 4 to 5 times 
larger ($\sim$0.5 dex) than the $\sigma$-values of $\sim$0.1 dex 
we compute for the curves-of-growth based on 
{\em computed} Weq-values ({\em middle panels}). 
The $\sigma$-values of these curves-of-growth 
are substantially larger due to random errors 
in the observed Weq-values compared to computed 
Weq-values in Fig. 4. The residual errors 
with respect to the mean are however too 
small to account for the large 
$\Delta$log(gf)-values (up to $\sim$2 dex) we 
measure for the majority of weak Fe~{\sc i} and 
Si~{\sc i} lines. It is of note that the 
$\sigma$-values of the curves-of-growth 
based on observed Weq-values are useful 
to compute {\em absolute} errors of log(gf)-values 
we iteratively measure with the detailed spectral 
synthesis method. The minimization of the 
standard deviation of the curves-of-growth 
is useful to test if stellar atmosphere models with  
for example depth-dependent microturbulence 
velocity or elemental abundance gradients 
can further improve the overall best fits to  
optical stellar spectra. However, more advanced radiative transfer 
modeling work (multi-D, non-LTE, etc.) to determine very 
accurate elemental abundance values from a mean curve-of-growth method 
should be based on carefully selected 
sets of spectral lines of which the atomic input data 
(e.g., log(gf)-values, energy levels, damping constants) 
are obtained from laboratory measurements.
However, laboratory measurements are currently not available 
for the majority of neutral optical lines of 
which we adjust log(gf)-values provided in atomic databases.
The atomic input data of the lines spectrum 
has to be predicted with advanced semi-empiric 
computations using specialized computer codes 
(i.e. HFR Cowan codes). 
Such computations of approximate log(gf)-values 
are very important to provide sufficient amounts
of line opacity for stellar spectral synthesis 
calculations required to develop realistic 
models of astrophysical stellar plasmas (e.g., 
thermodynamic models of stellar atmospheres).

A likely source for the systematic trends we find 
in Fe~{\sc i}, Ni~{\sc i}, and Si~{\sc i} lines 
therefore is the limited accuracy of predicted 
log(gf)-values derived from semi-empiric computations 
of weak lines. For a large number of weak 
lines only {\em predicted} restwavelengths, log(gf)-values, 
and damping constants are offered in atomic line 
databases. The accuracy of predicted log(gf)-values 
for weak lines of atoms with complex energy level structures 
is known to be limited. We find that they tend to be 
systematically over-estimated for many weak lines of  
neutral atoms. The Si~{\sc i} lines we adjust
belong to the same multiplets with lower energy levels 
($\chi_{\rm low}$) high in the atom (i.e. 4.95 eV and 5.619 eV). 
Semi-empiric computations of predicted log(gf)-values using
high energy levels are less accurate. In contrast,
we do not find a systematic trend in the $\Delta$s 
of 41 V~{\sc i} lines because many weak lines we adjust 
belong to multiplets with $\chi_{\rm low}$-values of 
0.27 - 0.30 eV and 1.05 - 1.21 eV, rather close to the 
ground level. These energy levels, rather low in the atom, are 
more accurately determined and yield on average predicted 
log(gf)-values that are more precise.  
For many high levels it is hard to compute the mixing 
in the eigenvectors correctly. Since the strengths of 
weak lines sensitively depend on the mixing the lines
have large errors. The problem of the limited accuracy of 
predicted atomic data for weak optical lines is also 
clear from the fact that we must remove 7 predicted 
Si~{\sc i} lines and 14 Fe~{\sc i} lines from the 
spectrum synthesis input lists because they are not 
observed in the Sun and other stars. 

\section{Conclusions}

We measure accurate log(gf)-values of 911 atomic 
absorption lines of neutral elements observed in 
high-resolution stellar spectra. We perform 
detailed synthesis calculations of the optical 
spectrum of the Sun, Procyon, and $\epsilon$ Eri 
observed with very large S/N ratios exceeding 2,000. 
We find systematic over-estimations in the log(gf)-values adopted 
from atomic databases for weak lines of iron-peak elements 
(such as Fe~{\sc i}, Ni~{\sc i}, Cr~{\sc i}, and Ti~{\sc i}), 
and of Si~{\sc i}. We demonstrate that the computed 
equivalent line widths strongly correlate
to observed values previously published for the lines in the
optical spectrum of the Sun. An average curve-of-growth 
analysis reveals that the errors of the log(gf)-values 
we measure for weak lines are smaller than the standard
mean error. The systematic trends we find for weak lines 
therefore cannot be attributed to large systematic errors in the 
detailed spectral line synthesis modeling method of the 
three stars. We attribute the remarkable systematic trends to the 
limited accuracy of predicted log(gf)-values of weak 
absorption lines having central depths below 15 \% 
currently offered in online atomic databases. The adjusted 
log(gf)-values we measure are available in the online 
SpectroWeb database, together with the observed and computed 
spectra and the central line depths and equivalent width values 
computed for the three stars. Further updates of the 
log(gf)-values and other atomic line data based on 
accurate laboratory measurements and advanced semi-empiric 
computations are urgently needed for reliable line identifications 
in the optical spectra of stars of all spectral subtypes. 

\section*{Acknowledgements}
The author acknowledges funding from the ESA/Belgian Federal 
Science Policy in the framework of the PRODEX programme 
(C90290). Mr. P. De Poorter and Mr. T. Hendrix are gratefully 
acknowledged for assistance with the spectral line measurements.
We thank the referees for serveral useful comments and questions.
The Hermes project is a collaboration between the KULeuven, 
the Universit\'{e} Libre de Bruxelles and the Royal Observatory of Belgium 
with contributions from the Observatoire de Gen\`{e}ve (Switzerland) 
and the Th\"{u}ringer Landessternwarte Tautenburg (Germany). 
Hermes is funded by the Fund for Scientific Research of Flanders 
(FWO) under the grant G.0472.04, from the Research Council of 
K.U.Leuven under grant GST-B4443, from the Fonds National de la 
Recherche Scientifique under contracts IISN4.4506.05 and 
FRFC 2.4533.09, and financial support from Lotto (2004) 
assigned to the Royal Observatory of Belgium.

\newpage
\begin{figure*}
\includegraphics[width=4.0in,angle=-90]{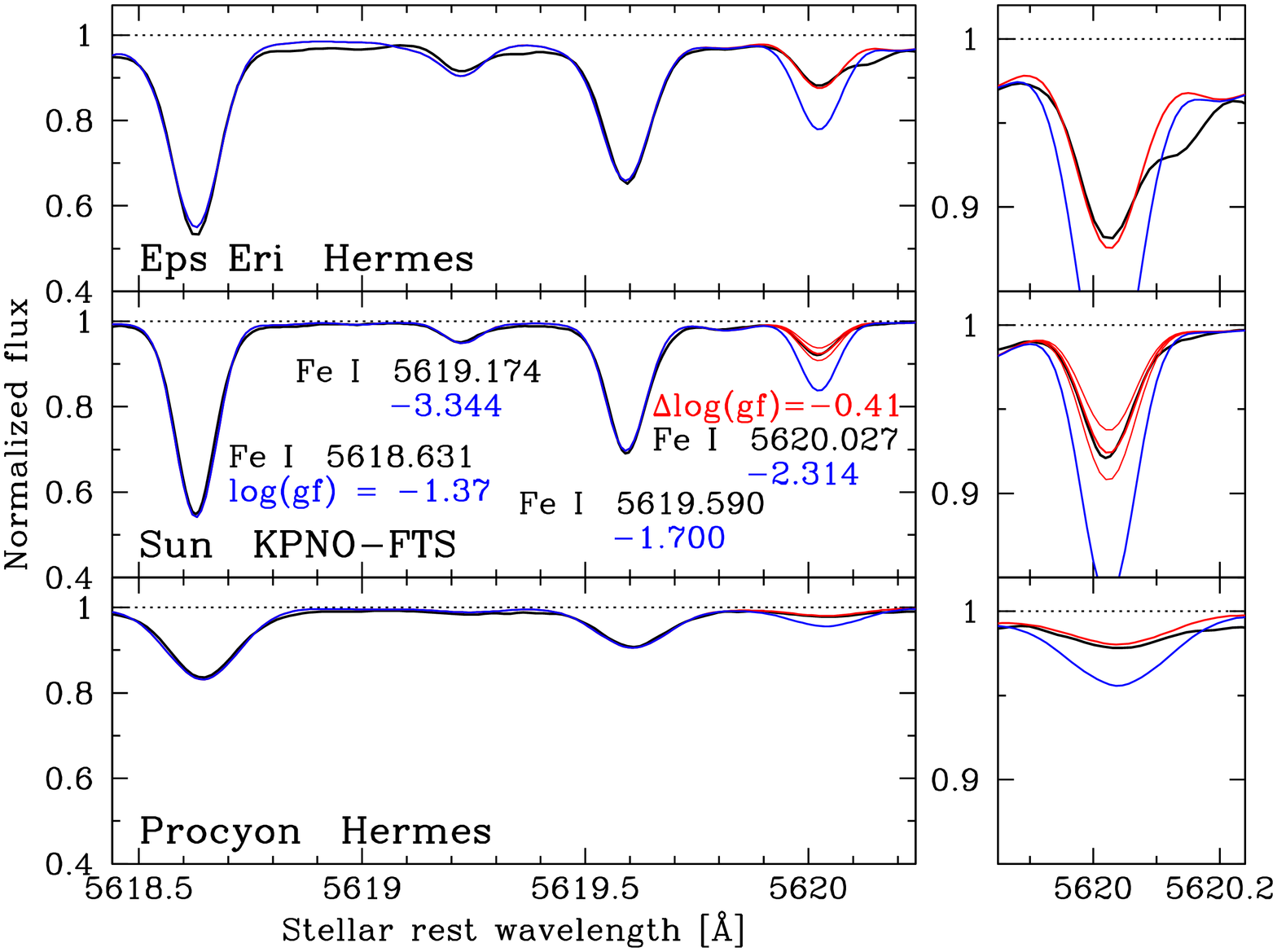}
\caption{
The solid drawn (black) lines show large signal-to-noise ratio 
flux normalized spectra of $\epsilon$ Eri ({\it top panels}), 
the Sun ({\it middle panels}), and Procyon ({\it bottom panels}),
compared to the best fit theoretical spectra ({\it solid blue lines}) 
calculated with (literature) log(gf)-values adopted from 
atomic databases ({\it marked blue numbers}). The depth of 
the weak Fe~{\sc i} line around 5620~\AA\, is computed too
strong compared to the observed line depth. A decrease 
of $\Delta$log(gf)=$-$0.41 yields the correct line depth 
in the three stars ({\it solid red lines}). 
The variation in the line depth of the best fit log(gf)-value
with $\pm$0.1 is also shown for the Sun ({\it thin drawn 
red lines in middle panels}). The right-hand panels show 
the fits to the Fe~{\sc i} $\lambda$5620 line in more detail.
}
\end{figure*}

\begin{figure*}
\includegraphics[width=4.0in,angle=-90]{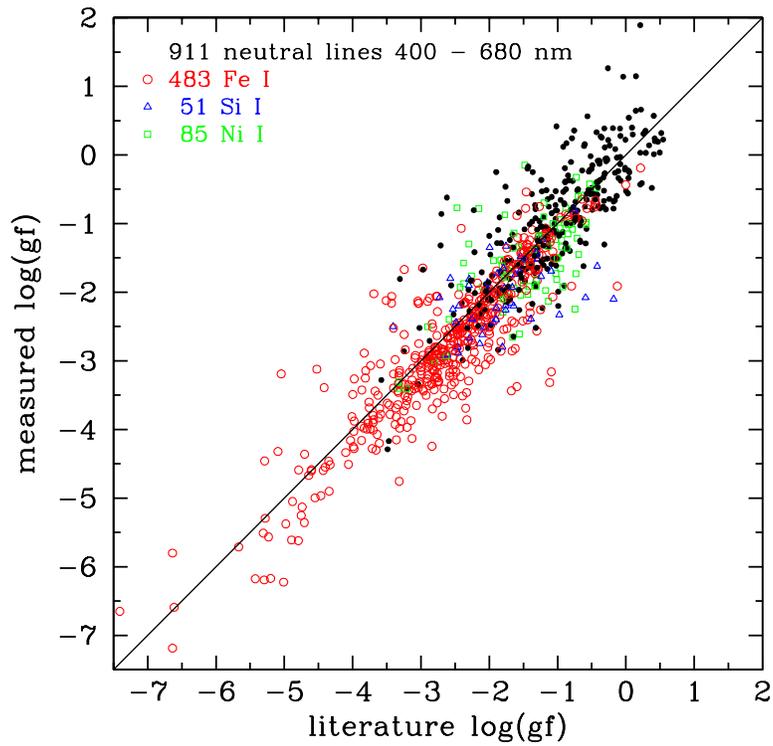}
\caption{ Comparison plot of 911 log(gf)-values 
provided in atomic databases (`literature log(gf)-values') 
and the log(gf)-values we measure with detailed spectrum 
synthesis calculations of the spectra of $\epsilon$ Eri, the Sun, 
and Procyon observed with very large S/N ratios between 400 nm and 
680 nm. We find a remarkable systematic trend of smaller measured 
log(gf)-values compared to the literature log(gf)-values towards 
the weaker lines ({\it open red circles mark neutral iron lines}), 
and of larger measured mean log(gf)-values towards the stronger 
lines of the sample ({\it solid black dots}) ({\it see text}). 
}
\end{figure*}

\begin{figure*}
\includegraphics[width=4.0in,angle=-90]{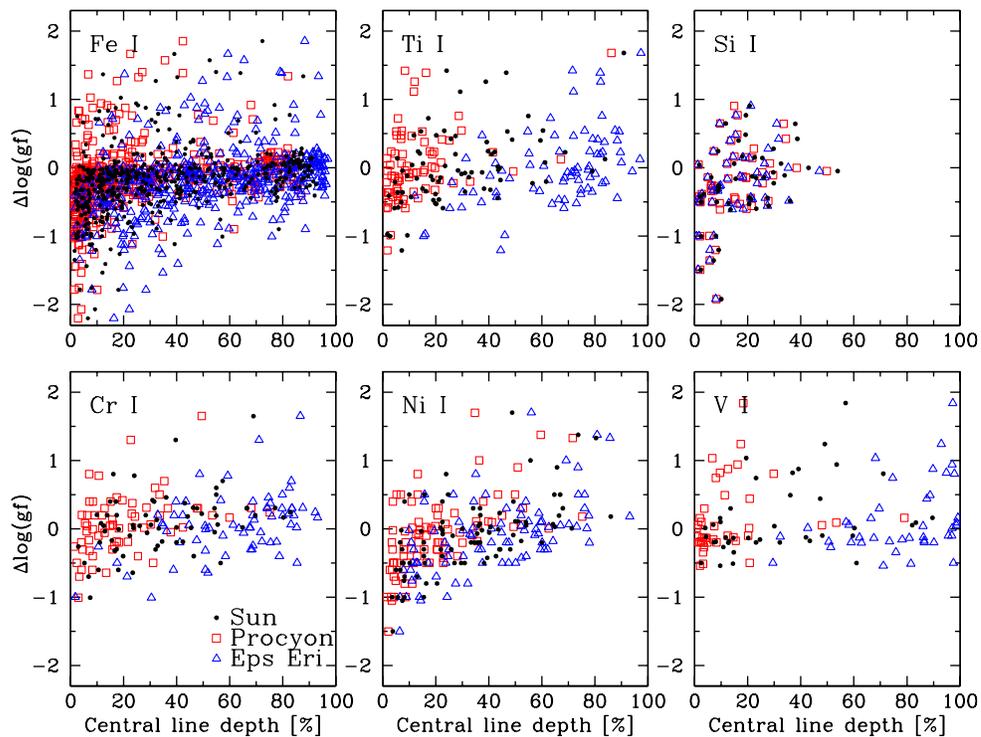}
\caption{ The panels show the $\Delta$log(gf)-values (the adjustment to 
the literature log(gf)-values to correctly fit the observed spectra)
with computed normalized central line depths in $\epsilon$ Eri 
({\it open triangle symbols}), the Sun ({\it solid black dots}), and
Procyon ({\it open red square symbols}). We find systematic trends of more 
negative $\Delta$log(gf)-values towards weak lines with central line
depths below 15\% in Fe~{\sc i}, Ni~{\sc i}, and Si~{\sc i}. 
The systematic trend is for example not observed in 41 lines of
V~{\sc i} ({\it lower right-hand panel}).} 
\end{figure*}

\begin{figure*}
\includegraphics[width=4.0in,angle=-90]{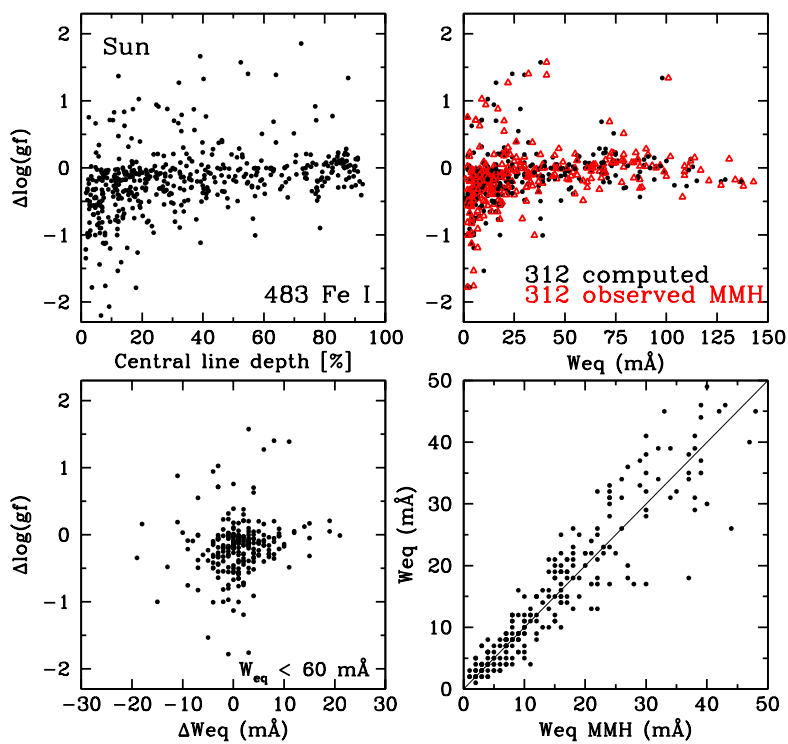}
\caption{ The $\Delta$log(gf)-values of 312 Fe~{\sc i} lines
having equivalent line widths observed by MMH in the Sun ({\it open red triangle symbols in upper right-hand panel}) are compared to Weq-values measured with spectral line synthesis calculations 
in this paper ({\it solid black dots}). The lower right-hand 
panel shows a strong correlation of the observed and computed Weq-values. The differences between the observed and computed Weq-values ($\Delta$Weq) are independent of the $\Delta$log(gf)-values ({\it lower left-hand panel}). 
It signals that the systematic trend observed in weak Fe~{\sc i} 
lines cannot be attributed to large systematic errors in the spectral
line synthesis calculations ({\it see text}).} 
\end{figure*}

\begin{figure*}
\includegraphics[width=4.2in,angle=-90]{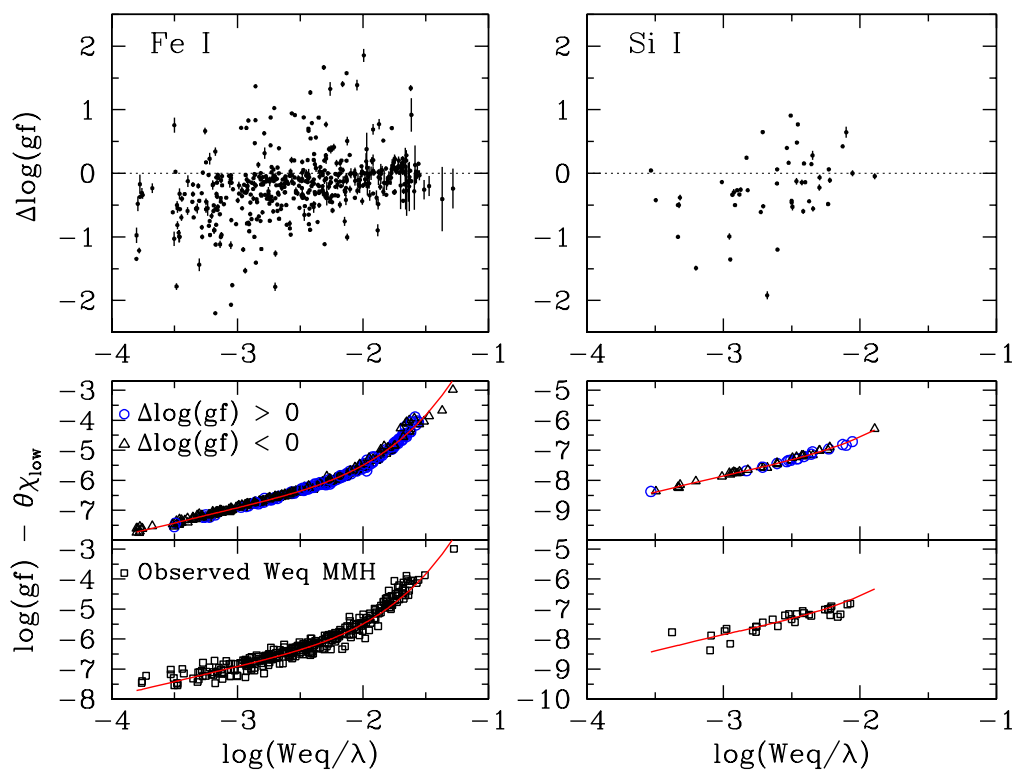}
\caption{ The upper panels show the $\Delta$log(gf)-values
with the reduced equivalent line widths (log(Weq/$\lambda$)) 
calculated in the Sun with best fit spectral synthesis modeling of 
483 Fe~{\sc i} ({\it left-hand panel}) and 51 Si~{\sc i} 
lines ({\it right-hand panel}). The middle panels 
show the corresponding average curves-of-growth 
({\it solid red lines}) used to compute relative
errors (marked in the upper panels) of the measured 
log(gf)-values. The lower panels show curves-of-growth
based on equivalent line width values observed by MMH
({\it see text}).
}
\end{figure*}

\end{document}